# Strategic Thinking for Sustainable Practice-Centered Computing

Myriam Lewkowicz, LIST3N/Tech-CICO, Université de Technologie de Troyes, France

This book was a trigger for me to reflect on all the projects we have conducted at the local, regional, national, and European levels around healthcare, addressing the social isolation of elderlies, the social support among caregivers, and the coordination issues of professionals willing to take care for patients at their home. Since Ina Wagner interviewed me in Paris, the situation evolved in France as a new healthcare law was voted that emphasizes the importance of cooperation among healthcare practitioners, the need for IT support, and the key role of local territories (which is not obvious in France, that is a very centralized country). The reflection that started with this book definitely helped me to make decisions in this evolving context, and I am grateful to Claudia Müller who, through an interview, facilitated me to realize what are, from my experience, the main issues that have to be dealt with when aiming at ensuring a sustainable practice-based computing approach. In this epilogue, I briefly list and illustrate these main issues, hoping they could support other sustainable experiences.

## Following topics

One of the most important and obvious thing to do is to identify research questions that are independent of a case, and to ensure that projects can ensure a follow-up of these research questions. For instance, I studied knotworking (Engeström, 2008) with Khuloud Abou Amsha in different settings. In PiCADO, a three years long inter-ministerial funded project, we studied the collaborative practices of self-employed healthcare professionals (self-employed workers in charge of the economics of their practice: general practitioners, nurses, speech-language pathologists, physiotherapists, occupational therapists, dieticians, etc. ) taking care of patients at home (Abou Amsha & Lewkowicz, 2018). In HADex, a one year long regionally funded project, we studied the coordination among salaried healthcare professionals of the "hospital-at-home" organization. Finally, in CALIPSO, a two years long regionally funded project, we studied the coordination between the different healthcare actors for the whole trajectory of a patient, from hospital to home, and vice versa, addressing the sharing of information and the communication between two very different healthcare systems (at least in France): in the city and at the hospital (Berthou, tbp.). By adopting the design case study framework (Wulf et al., 2011), we were able to develop a conceptual reflection on knotworking and how to design to support it (Abou Amsha et al., 2021), reflection that led us to develop a discourse towards the local and regional healthcare actors in France so that our research could impact local communities.

## Getting political support

Having an impact on local communities is one important aspect of socio-informatics that we embrace, ensuring that the research questions that we tackle are those that are "important, meaningful, and impactful for the communities" with whom we are working (Hayes, 2018 p. 308). Created by decree of September 14, 1994, the Troyes University of Technology was born out of a strong political will. The general council of Aube - and more precisely its president, Senator Philippe Adnot - bet on technological development and innovation to save the economy of the department, highly industrialized (fourth French department) but suffering from the crisis in the textile and the mechanical industries (Moraux & Balme, 2007). This favorable political context encouraged us to interact regularly with the Senator, who then became aware of the research we were conducting, and convinced by the positive impact it could have for citizens, local companies, students, and research groups. He and his team then regularly invited us to lunches at the Senat to introduce us to major actors of companies in the healthcare sector (insurances, IT, transportation), envisioning our research as a way to attract diverse economic players in the department. He also invited us to national events, and discussed policies debated in the French Senat. This political support has definitely opened us prospects and given me a complementary vision that was very useful in making strategic decisions.

## Being flexible

Something that I have learnt after fifteen years of research within healthcare is that flexibility, in the sense of being able to adapt and to change strategies as the context evolves, is another important aspect of sustainability. I will illustrate this aspect with our journey through a major evolution of our ecosystem.

When we defined the CALIPSO project as a follow-up project of PiCADO, our objective was to build up upon CARE (Amsha & Lewkowicz, 2015), the prototype that was developed to support knotworking. For so doing, we involved a local (based in Troyes) software company that was used to work with the local hospital in the project. The goal was both to deliver a robust product, and to ensure its maintenance, which would ensure the sustainability of the whole approach. However, this initial plan was disrupted.

The disruption started with the French territorial reform in 2016 that reduced the number of regions from 22 to 13. Before this reform, our research group was one of the few working on coordination among healthcare practitioners in our region (Champagne-Ardenne). The new region, called Grand-Est, includes Lorraine and Alsace, two previous regions on their own, that contain major research centers, big hospitals, and well organized associations. A regional agency, named Pulsy, was created to be the preferred operator of the Grand-Est Regional Health Agency for the development and implementation of the regional e-health strategy. It aims first at supporting and promoting the use of digital health services in the territories to health professionals, health establishments, social and medico-social structures, and users. Its objective is also to facilitate the sharing and exchange of data, in a standardized and secure framework, to coordinate care and life paths, to support its members in the implementation of regulatory obligations and standards of good practice, and to promote innovation and territorial initiatives in the field of e-health (www.pulsy.fr). Pulsy therefore proposes a catalog of systems and related services that interferes with potential local developments. Indeed, in the French healthcare system, a hospital cannot fully decide which information systems they may use; if they decide to use a system that is not "prescribed" by its Regional Health Agency, they are not financially supported for its acquisition, its customization, nor its deployment.

We were then confronted to a difficult situation: should we go on working with the local software company, but end up with a system that fits our research results but that would not be deployed by the local hospital anymore… We faced a dilemma, as supporting a practice-centered approach at the regional level would mean losing the system that we contributed developing during two design case studies. Keeping in mind a long-term perspective, we decided not to fight against the regional e-health agency and to accompany the national strategy, applied at the local level. The decision was not easy, as the software company was also at risk of losing an important market, leading to economic lay-off. The system that was supposed to be developed during CALIPSO was then never finished, but we are now involved with the local hospital in the deployment of Parcéo, a system supported by Pulsy that calls it "the whats app for the nurses and the doctors". By making the point that we have an expertise in this domain, we managed to be integrated into the project and to have a say in the customization and the deployment process. A PhD student from our research group (paid by the university) is now following the whole process, aiming at pushing in favor of a practice-centered point of view.

## Ensuring independence through diverse sources of funding

Another important aspect of sustainability is to balance the financial and the intellectual independence. After fifteen years of responding to calls for projects, I have now learnt to adopt a portfolio approach, mixing short and longer projects, funded by companies or administrations (hospital, healthcare agencies), and by public authorities. Indeed, public funding ensures intellectual independence but also means more bureaucracy, more time spent on deliverables (and less on writing manuscripts), whereas direct contracting with companies or administration means money that can be spent exactly as we want (without having to respect a budget plan defined three or four years ago). I adopt the same approach for funding PhD; it is interesting to have a company funding a PhD (in France there is a financial incentive scheme) as the company becomes interested in the success of the research, the PhD student gains an employment experience, and the research group is financially supported for the supervision of the student. On the other hand, it becomes more difficult to ensure fully critical thinking in this framework. Therefore, it is important to decide which topic to address under which framework.

For instance, in the new context, it is crucial to start interacting closely with Pulsy, the new e-health regional agency presented above. I have then joined forces with a colleague based in Strasbourg (where the headquarters of Pulsy are) and negotiated a PhD there, who is going to work on the regional version of the national e-health strategy. However, as mentioned above, when following the deployment of Parcéo (the system supposed to support the cooperation among all the care actors), it is important to ensure that a publicly funded PhD student is working on the case. It guarantees that we will be able to raise concerns about the system and its deployment and publish about that.

## Postponing and pausing activities

Even if following topics is an important factor for sustainability, that we mentioned first, it goes with being able to postpone and pause activities. Indeed, what is important is to keep track of open research questions, and to be opportunistic when deciding to address them.

For instance, we started to work on teleconsultation in 2015, as both the head of the emergency medical dispatch center in Troyes, and the head of a company offering a telemedicine toolkit contacted us to assess the deployment of this telemedicine toolkit in ten nursing homes. While conducting this assessment, we got interesting findings (Gaglio et al., 2016), but did not have time to explore them in detail, partly because our sociologist colleague who led the study, Gérald Gaglio, got a full professor position in another university. However,

I kept that in mind, and when in 2019, I got some funding through one of the sponsors of the SilverTech Chair of our university, I took this opportunity to reopen the topic; I went back to the doctor in charge of putting in place teleconsultation between the local hospital and nursing homes. I offered him to welcome a master student to study what they have put in place, which he accepted. The study (Cormi et al., 2020) really interested the hospital, which Director decided to hire the student for a PhD.

## Building a team

The last but not least sustainability aspect is related to hiring the right people and networking them with the local and regional institutions. Indeed, what was key in the development of our research in the last ten years was the possibility to hire two assistant professors, Matthieu Tixier first, in 2013, and Khuloud Abou Amsha four years after. Historically, and until shortly, they both would work with me, but not with each other. So, when I had the occasion to reopen the teleconsultation topic as described above, I offered them to supervise the student together. It went well, and I supported them by interacting with the head of the hospital when they were defining the PhD project for the master student that they are now supervising together.

I envisioned my role as a mentor, positioning them on some topics, supporting them finding the funding, and let them grow, while being present if they need any advice. By this strategy, I can also share the responsibility with Khuloud and Matthieu and I do not have to carry everything by myself; having more people involved at the university is definitely a matter of sustainability.

We have also recently hired an assistant professor in management science, Loubna Echajari, conducting research related to crisis management in industry. As the pandemic started, she asked if I could put her in contact with the hospital. I did so, asking them if they were interested by a study that our research group would fund. They accepted, her presence and her work (Sanchez et al., 2020) were really appreciated, which played an important role in how the hospital is seeing our research group as highly pertinent, compared to the other research groups from our university they are working with (related to optimization and data management).

## Conclusion

Finally, looking back at fifteen years of research applied in the healthcare domain conducted at the Troyes University of Technology, trying to have a positive impact on patients, their informal caregivers, and the professionals, I would say that the strategy that has been developed and followed is aligned on the strategy that was envisioned in 1994, when the University was created thanks to the strong political will of Senator Philippe Adnot. All the activities described in chapter 5, analyzed in chapter 11, and reflected in this epilogue demonstrate some of the "lived practices" in executing this overall strategy. Through these activities, we took part in the development of the local associations and in the positioning of the local hospital, and there is more to come.

Indeed, last year, the department has been chosen as one of the five departments in France (that counts 101 departments) to experiment what has been called "population responsibility". It corresponds to the "triple aim" approach put in place in North America (Berwick et al., 2008): a better care experience for the patient, a better health for the population, and a lower cost for society. It also involves going beyond city-hospital oppositions, and focusing on prevention. By taking part in this initiative that entails ambitious improvement at all levels of the healthcare system, we are pursuing our strategy in favor of sustainable practice-centered computing that, as we have showed, must be envisioned as a local cooperative endeavor, with all its struggles, but also new kinds of opportunities.

## References

Abou Amsha, K., Bossen, C., & Grönvall, E. (2021, tbp). Computer-Supported Knotworking Design guidelines based on two case studies from the healthcare domain in Europe. *PACM HCI*.

Abou Amsha, K., & Lewkowicz, M. (2018). Chapter 3—Supporting Collaboration to Preserve the Quality of Life of Patients at Home—A Design Case Study. In M. S. Ackerman, S. P. Goggins, T. Herrmann, M. Prilla, & C. Stary (Éds.), *Designing Healthcare That Works* (p. 39-57). Academic Press. https://doi.org/10.1016/B978-0-12-812583-0.00003-1

Amsha, K. A., & Lewkowicz, M. (2015). CARE : An Application to Support the Collective Management of Patients at Home. *2015 International Conference on Computational Science and Computational Intelligence (CSCI)*, 743-748. https://doi.org/10.1109/CSCI.2015.22

Berthou, V. (tbp.). Un carnet de santé pour mieux coordonner ? Une analyse sociologique de la dynamique organisationnelle du projet Calipso. In *La révolution digitale en santé—Comment innover et agir en faveur de la mutation du système de santé ?* (Grenier C., Rizoulières R., Béranger J. (Eds.)). ISTE.

Berwick, D. M., Nolan, T. W., & Whittington, J. (2008). The Triple Aim : Care, Health, And Cost. *Health Affairs*, *27*(3), 759-769. https://doi.org/10.1377/hlthaff.27.3.759

Cormi, C., Abou Amsha, K., Tixier, M., & Lewkowicz, M. (2020). *How the local domestication of a teleconsultation solution is influenced by the adoption of a national policy?* https://doi.org/10.18420/ecscw2020_ep06

Gaglio, G., Lewkowicz, M., & Tixier, M. (2016). « It is Not Because You Have Tools that You Must use Them » : The Difficult Domestication of a Telemedicine Toolkit to Manage Emergencies in Nursing Homes. *Proceedings of the 19th International Conference on Supporting Group Work*, 223-233. https://doi.org/10.1145/2957276.2957288

Hayes, G. R. (2018). Design, Action, and Practice : Three Branches of the Same Tree. In *Socio-Informatics*. Oxford University Press. https://oxford.universitypressscholarship.com/view/10.1093/oso/9780198733249.001.0001/oso-9780198733249-chapter-10

Moraux, M.-F., & Balme, P. (2007). *Université de Technologie de Troyes* (n° 2007-003). Inspection générale de l'administration de l'Éducation nationale et de la Recherche. https://www.vie-publique.fr/sites/default/files/rapport/pdf/174000079.pdf

Sanchez, S., Echajari, L., Friot-Guichard, V., Blua, P., & Laplanche, D. (2020, octobre 13). L'évolution des routines comme levier de l'innovation organisationnelle : Le cas du Département d'Information Médicale du Centre Hospitalier de Troyes face à la crise Covid-19. *L'innovation organisationnelle et managériale en santé Cadre théorique, analyse d'impact sur les pratiques professionnelles et l'organisation des soins Attentes particulières sur les retours d'expérience de la gestion*. 8e Congrès Aramos, Ecole polytechnique Paris.

In Section IV, Epilogue of: SIMONE, Carla, WAGNER, Ina, MÜLLER, Claudia, *et al. Future-proofing: Making Practice-Based IT Design Sustainable*. Oxford University Press, 2022, pp.341-347. https://global.oup.com/academic/product/future-proofing-9780198862505?q=Future-proofing:%20Making%20Practice-Based%20IT%20Design%20Sustainable&cc=fr&lang=en

Wulf, V., Rohde, M., Pipek, V., & Stevens, G. (2011). Engaging with practices : Design case studies as a research framework in CSCW. *Proceedings of the ACM 2011 Conference on Computer Supported Cooperative Work - CSCW '11*, 505. https://doi.org/10.1145/1958824.1958902